\documentclass[runningheads]{llncs}
\usepackage{graphicx}
\usepackage{multicol}
\usepackage{graphicx}
\usepackage{wrapfig,lipsum,booktabs}
\usepackage{multirow}
\usepackage{tcolorbox}
\usepackage{xcolor}
\usepackage{colortbl,hhline}
\usepackage[export]{adjustbox}
\usepackage{adjustbox}
\usepackage{comment}
\usepackage{paralist}
\usepackage[textsize=scriptsize]{todonotes}

\begin{document}
\title{The rise and rise of interdisciplinary research: Understanding the interaction dynamics of three major fields -- Physics, Mathematics \& Computer Science}%\thanks{Supported by organization x.}}
\titlerunning{The rise and rise of interdisciplinary research}
\author{Rima Hazra\inst{1} \and
Mayank Singh \inst{2} \and
Pawan Goyal\inst{1}\and
Bibhas Adhikari\inst{1} \and
Animesh Mukherjee\inst{1}}
\authorrunning{Hazra et al.}
\institute{Indian Institute of Technology, Kharagpur, India\\
\email{to\_rima@iitkgp.ac.in,\{pawang,animeshm\}@cse.iitkgp.ac.in,bibhas@maths.iitkgp.ac.in} \and
Indian Institute of Technology, Gandhinagar, India\\
\email{singh.mayank@iitgn.ac.in}}
\maketitle              % typeset the header of the contribution
\begin{abstract}
The distinction between sciences is becoming increasingly more artificial -- an approach from one area can be easily applied to the other. More exciting research nowadays is happening perhaps at the interfaces of disciplines like Physics, Mathematics and Computer Science. How do these interfaces emerge and interact? For instance, is there a specific pattern in which these fields cite each other? In this article, we investigate a collection of more than 1.2 million papers from three different scientific disciplines -- Physics, Mathematics, and Computer Science. We show how over a timescale the citation patterns from the core science fields (Physics, Mathematics) to the applied and fast-growing field of Computer Science have drastically increased.  Further, we observe how certain subfields in these disciplines are shrinking while others are becoming tremendously popular. For instance, an intriguing observation is that citations from Mathematics to the subfield of machine learning in Computer Science in recent times are exponentially increasing. 

\keywords{Interdisciplinarity  \and Computer Science \and  Mathematics \and Temporal Bucket Signatures.}
\end{abstract}

\section{Introduction}
\label{sec:intro}
Science is built upon previous knowledge, which spans over ideas and concepts drawn from multiple disciplines. The availability of scientific literature from different disciplines of research is vastly expanding due to the advancement in the Internet infrastructure. Several recent studies show how these disciplines interact with each other. Organization for Economic Cooperation and Development (OECD, 1998)~\cite{OECD:1998,Morillo:2003} defines three classes of research based on several levels of interactions among disciplines: (i) multidisciplinary, (ii) interdisciplinary, and (iii) transdisciplinary research. In multidisciplinary research paradigm, researchers from different fields collaborate together, but each of them confines their research to their disciplinary boundaries and exploit their own domain knowledge to address the problem. In contrast, in interdisciplinary research, researchers integrate concepts of different fields to solve their domain problems. The transdisciplinary research adds another dimension in which researchers create an intellectual group beyond their field. 

The current work focuses on interdisciplinary research. 
The main objective of interdisciplinary research is to improve fundamental approaches or to solve problems whose solutions are not in the scope of a single field of research practice~\cite{Porter:2009}. Interdisciplinary research is a common practice in science since early decades. Recent studies show how science is becoming highly interdisciplinary in the last few decades~\cite{Morillo:2003}. This work presents a thorough analysis of citation interactions to demonstrate interdisciplinarity among scientific disciplines. Citation interactions are represented by bibliographic relations such as ``\textit{who cites whom}'', ``\textit{when one cites other}'', etc. We focus on ``\textit{who cites whom}'' relationships to quantify interdisciplinarity. Here, ``who'' represents citing field and ``whom'' refers to cited field. As a case study, for the first time, we conduct empirical experiments on three research fields Computer Science ($CS$), Physics ($PHY$), and Mathematics ($MA$) as these fields are closely interacting among themselves and exchanging their domain knowledge to address critical problems within their domains. Overall, the main objectives of this work are twofold: 
\begin{enumerate}
    \item Investigating patterns of citations across research fields.
    \item Thorough analysis of citation interactions leading to the interdisciplinarity of research fields.
    %\item Modeling the emergence of interdisciplinarity through network growth models.
\end{enumerate}

\begin{table}[t]
\begin{minipage}[b]{0.45\linewidth}
\caption{Dataset description.}\label{tab:dataset}
\scriptsize
\begin{tabular}{|c|c|c|c|} \hline
{\bf Sl. No} & {\bf Fields} & {\bf \# papers} & {\bf \# subfields} \\ \hline
1 & Computer Science & 1,41,662 & 40\\ \hline
2 & Mathematics & 2,84,540 & 33 \\ \hline
3 & Physics & 8,04,360 & 50 \\ \hline
%\bottomrule
\end{tabular}
%\end{adjustbox}
\end{minipage}
\hspace{1.5cm}
%\end{wraptable}
%\begin{wraptable}{r}{4.5cm}%[tbhp]
%\centering
\begin{minipage}[b]{0.45\linewidth}
\caption{Filtered datasets.}\label{tab:filtered_dataset}
%\begin{adjustbox}{width=0.3\textwidth}
\scriptsize
\begin{tabular}{|c|c|c|c|} \hline
\textbf{Time period} & \multicolumn{3}{c|}{\bf \# papers per field} \\ \cline{2-4}
   & ${\bf MA}$& ${\bf PHY}$& ${\bf CS}$ \\ \hline
 {\bf B1} (1995-1999)&  726& 7553&86\\ \hline
{\bf B2} (2000-2004) & 2452 & 13912& 104\\ \hline
{\bf B3} (2005-2009)& 5913&19748& 648\\ \hline
{\bf B4} (2010-2014) &5535 &12609 &4006\\ \hline
{\bf B5} (2015-2017)&10449&29896& 3186\\ \hline
\end{tabular}
%\end{adjustbox}
\end{minipage}
\vspace{-0.5cm}
\end{table}

\section{Datasets}
\label{sec:dataset}
We crawled \textit{arXiv}\footnote{www.arxiv.org}, one of the well-known pre-print repositories and collected all the research articles --- metadata as well as \LaTeX\ source code --- submitted between 1990--2017. It contains more than 1.2 million articles published in nine major fields. Each research field is further sub-divided into several subfields. For our experiments, we select the three major fields -- Computer Science ($CS$), Mathematics ($MA$), and Physics ($PHY$). The total number of papers in each field and the respective number of subfields is noted in Table~\ref{tab:dataset}. Next, the citation network among papers is constructed by parsing the references present in ``.bbl'' files. For each candidate paper, we only extract those referenced papers that are available in \textit{arXiv} by matching title string. In our experiments, we consider only those papers which have at least five extracted references.

\section{Empirical analysis}
\label{sec:emp_analysis}

We conduct an in-depth temporal analysis of citation interactions among the three disciplines. We group citation interactions into multiple buckets based on the publication year of the citing paper. We report results for bucket size of five years\footnote{Our results hold for other bucket sizes also. As, papers between 1990--1995 are very less in number, we start buckets from the year 1995.}.  We, divide the entire dataset (see Table~\ref{tab:filtered_dataset}) into five buckets (i) \textbf{B1} (1995--1999), (ii) \textbf{B2} (2000--2004), (iii) \textbf{B3} (2005--2009), (iv) \textbf{B4} (2010--2014), and (v) \textbf{B5} (2015--2017). 
The total number of papers belonging to each time period, and each major field are noted in Table~\ref{tab:filtered_dataset}. Citations gained from articles published within the same field are termed as \textit{self-field citations}. Incoming citations from other fields are termed as \textit{non self-field citations}. We observe, empirically, that the proportion of self-field citations is significantly higher than non self-field citations. We, therefore, study citation interactions at two levels -- (i) field and (ii) subfield level. Subfield level citation interactions present a more in-depth understanding of the interdisciplinary nature across these fields. 

\begin{figure*}[t]
\centering
\setlength\tabcolsep{-1pt}
\renewcommand{\arraystretch}{0.8}%
\begin{tabular}{@{}c@{}c@{}c@{}}
  \includegraphics[width=.33\hsize]{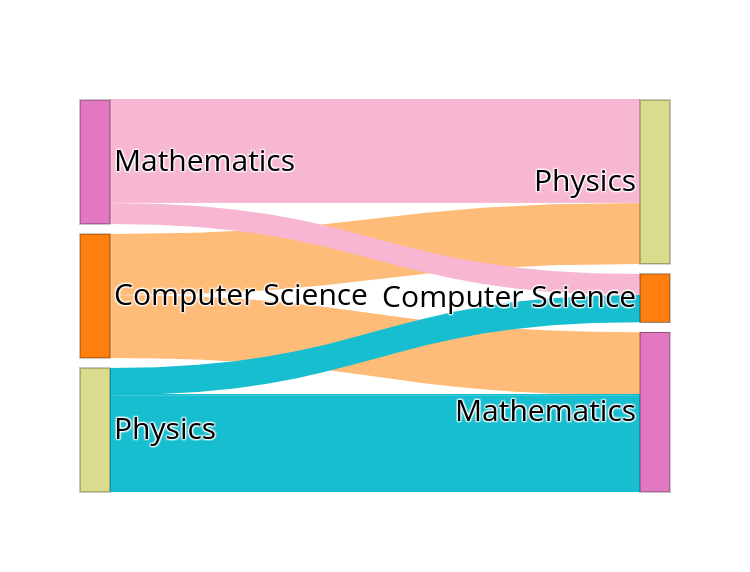} &
  \includegraphics[width=.33\hsize]{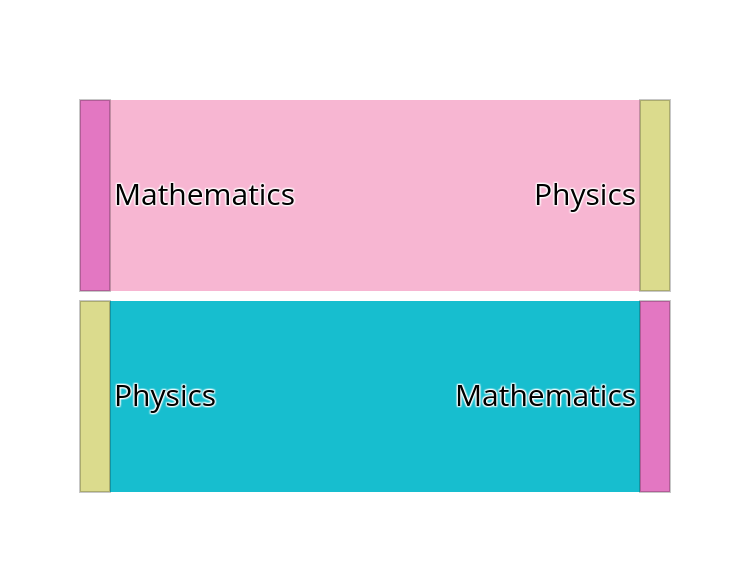}& 
  \includegraphics[width=.33\hsize]{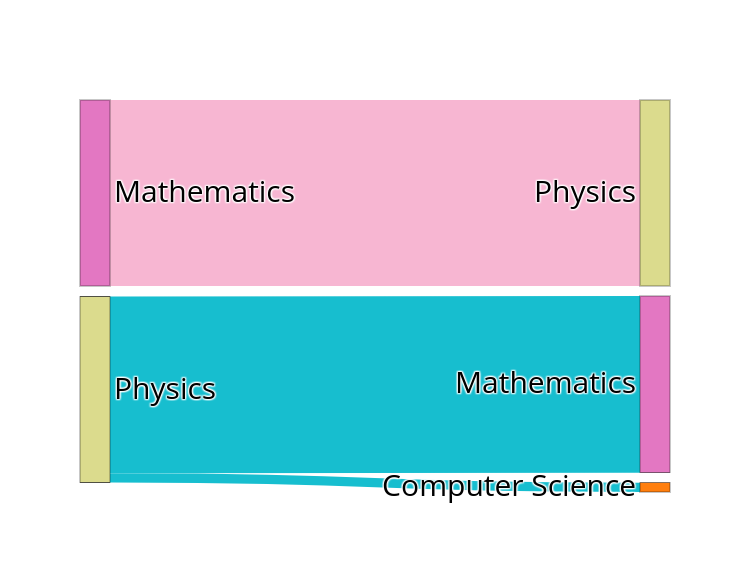}\\
  (a) Overall & (b) \textbf{B1} (1995--1999) & (c) \textbf{B2} (2000--2004)\\
  \includegraphics[width=.33\hsize]{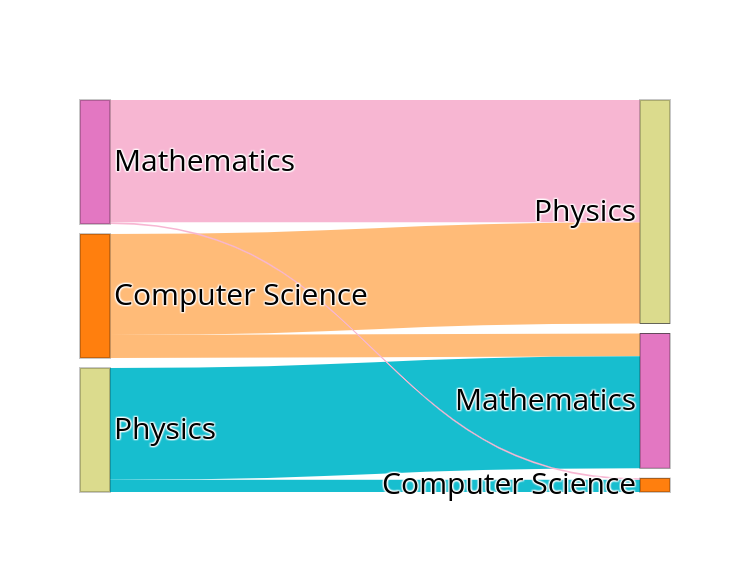} &
  \includegraphics[width=.33\hsize]{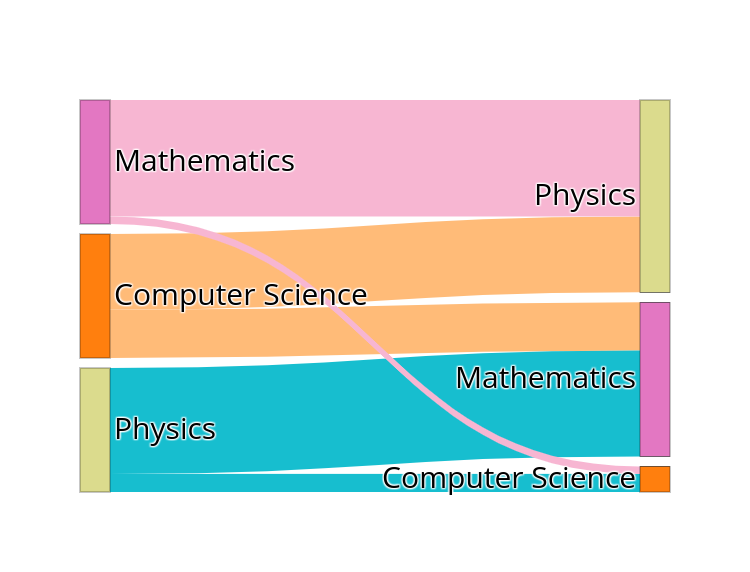}& 
  \includegraphics[width=.33\hsize]{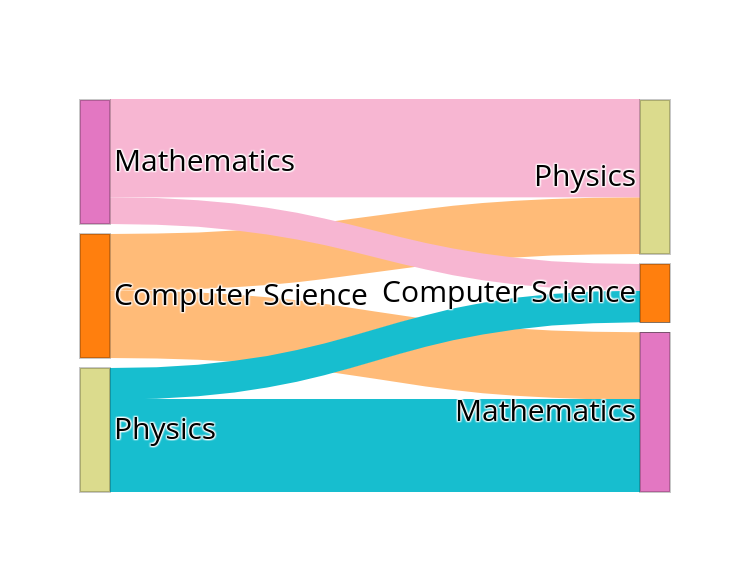}\\
  (d) \textbf{B3} (2005--2009)& (e) \textbf{B4} (2010--2014) & (f) \textbf{B5} (2015--2017)\\
\end{tabular}
\caption{Citation flow among three fields of science, Computer Science, Physics, and Mathematics over the (a) entire time-period, and (b-f) five temporal buckets.}
\label{fig:fraction_non_self}
\vspace{-0.5cm}
\end{figure*}

Sankey diagram is a graphical representation of flow from left to right in which width of the link is proportional to the amount of flow. In Figure \ref{fig:fraction_non_self}, we have shown the citation flow among $CS$, $MA$ and $PHY$. The leftmost nodes are represented as the source and the rightmost nodes are represented as the target. In our case, leftmost nodes are denoted as ``citer" field and rightmost nodes are denoted as ``cited" field. The link between source to target denotes the citation flow from the citer field to the cited field. 

\noindent\textbf{Observations}:
Figure~\ref{fig:fraction_non_self} shows citation flow among three fields. The temporal study uncovers several interesting observations. During initial time-periods, $CS$ was poorly cited by the other two fields (with no citation from $MA$ and $PHY$ in \textbf{B1}). All of the non self-field citations from $PHY$ went to $MA$ and vice-versa. However, in later time-periods, $CS$ started receiving attention from both $PHY$ and $MA$. It is also clearly evident that in the initial time-periods, $PHY$ was more cited by $CS$ than $MA$, however, the trends are reversed in the later time-periods. Note that, here, we do not consider the flow of self-field citations. We observe similar self-field citation flow trends for each field. We next discuss bucket-wise observations:

\noindent\textbf{B1} (1995--1999): During this time-period, $MA$ entirely cites $PHY$ and vice-versa. We do not observe in-/outflow of citations to/from $CS$.

\noindent\textbf{B2} (2000--2004): During this time-period, a marginal number of citations flow to $CS$ from $PHY$. $MA$, still, only cites $PHY$. Here, we do not observe outflow of citations from $CS$.

\noindent\textbf{B3} (2005--2009): During this time-period, $CS$ has started citing both the other fields; $CS$ seems to have been drawing more ideas from $PHY$ than $MA$ in this period. In contrast to the previous bucket, the number of citations from $PHY$ to $CS$ has increased.

\noindent\textbf{B4} (2010-2014): Interestingly, in this time span, we witness a complete shift in the citation patterns received by the $CS$ papers. In particular, $CS$ seems to have started receiving citations from $MA$.  

\noindent\textbf{B5} (2015-2017): $PHY$ and $MA$ both seem to be equally citing $CS$ papers in this span. We posit that this interesting trend could be mostly attributed to the newly emerging topics like {\em Deep Learning, Machine Learning, Statistical Natural Language Processing}, etc. These topics significantly borrow many ideas from Mathematics.

%Fraction of citation goes from one field to another field over years.
\begin{figure*}[!b]
\vspace{-0.5cm}
\centering
\includegraphics[width = 1\hsize]{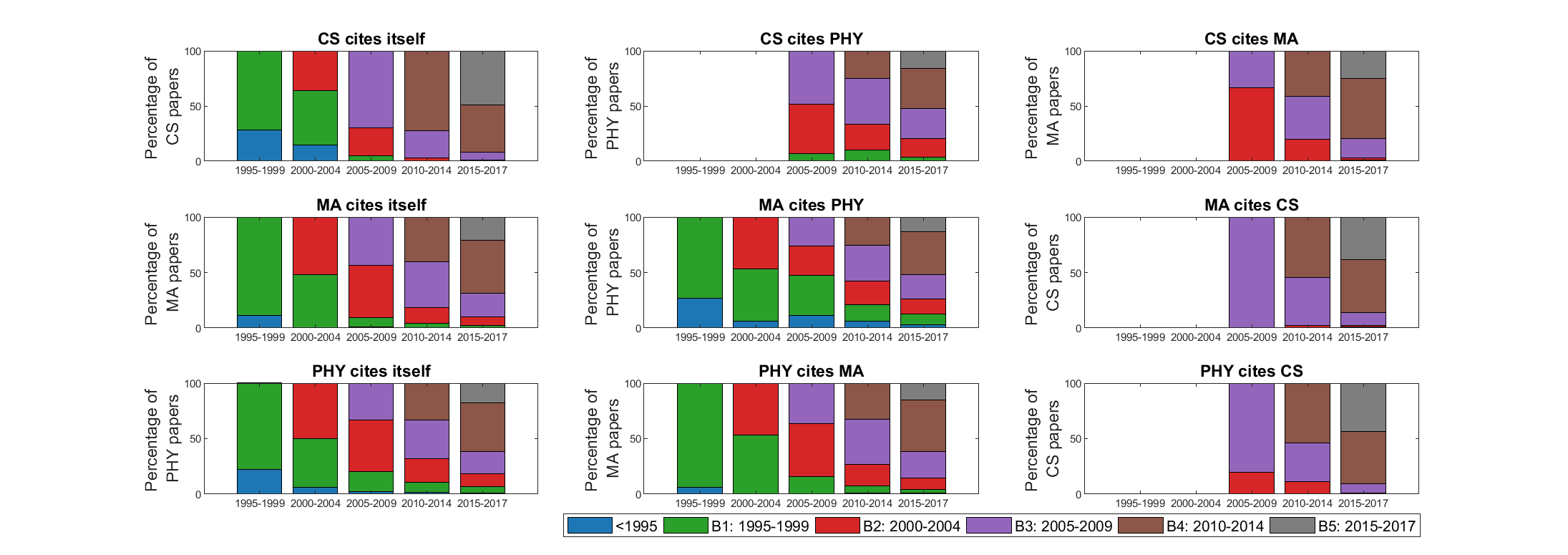}
\caption{Fraction of citation going from one field to another field over the years.}
\label{fig:3field_temporal}

\end{figure*}

Similar to Sankey diagrams, temporal bucket signatures (TBS)~\cite{Singh:2017} present a novel visualization technique to understand the temporal citation interactions. TBS refers to a stacked histogram of the relative age of target papers cited in a source paper. Suppose we collect citation links into fixed-size buckets of temporal width $T$ (e.g., $T=5$ years). We partition the entire article set into these fixed buckets based on publication year. For each bucket, we compute the fraction of papers of other buckets that are cited by the papers of the current bucket. As self-field citations are significantly larger in number, we create TBS for self-field citations and non self-field citations separately.  In the case of self-field citation TBS, we analyze how papers of the same field in older buckets receive citations from the current bucket papers. In the case of non self-field citation TBS, we observe how papers belonging to other fields in older buckets receive citations from current bucket papers.

\noindent\textbf{Observations}: 
%Next, we analyze pairwise citation interactions among fields over the years. 
%We leverage \textit{temporal bucket signatures} (TBS)~\cite{Singh:2017} to visualize proportional changes in field-wise inflow of citations across timescales. 
Figure~\ref{fig:3field_temporal} shows proportional citation flow from each field to other fields in different temporal buckets. $CS$, in contrast to $MA$ and $PHY$ cites current bucket papers more than the older bucket papers of its own field (evident from the higher proportion of the top segment in each temporal bucket). This observation reconfirms the common intuition that ``CS is fast growing field''. In contrast, $CS$  tends to cite older papers from the other two fields -- $MA$ and $PHY$ (denoted by a lower proportion of top-most segment in each temporal bucket). $MA$ and $PHY$ predominantly cite older papers of each other and recent papers from $CS$. 

\if{0}
\begin{table}[b]
\vspace{-0.5cm}
\centering
\caption{$p$-values for different citation trends, self-field citation, overall non self-field citations and pairwise non self-field citations.}\label{tab:sig_test}
%\scalebox{0.68}{
%\begin{adjustbox}{width=0.45\textwidth}
\tiny
\begin{tabular}{|c|c|c|c|c|c|c|} \hline
\multirow{2}{*}{\bf Citation} & \multicolumn{6}{c|}{\bf Year buckets} \\ \cline{2-7}
 & Overall & {\bf B1} & {\bf B2} &{\bf B3} & {\bf B4} & {\bf B5} \\ \hline
 $CS$ cites self field & 0.076 &    0.382 &    0.081 &    0.001 &    0.022 &    0.056 \\
 $MA$ cites self field &&&&&& \\ \hline
 $CS$ cites self field & 0.006    &0.083    &0.007    &0.000    &0.002 &0.010 \\
 $PHY$ cites self field &&&&&& \\ \hline
 $MA$ cites self field & 0.023 & 0.080 & 0.008 & 0.000 & 0.004 & 0.020 \\ 
 $PHY$ cites self field &&&&&& \\ \hline\hline
 $CS$ cites non self fields & 0.140 & 0.073 & 0.041 & 0.027 & 0.046 & 0.094 \\
 $MA$ cites non self fields &&&&&& \\ \hline
 $CS$ cites non self fields & 0.156 & 0.155 & 0.078 & 0.009 & 0.029 & 0.130 \\
 $PHY$ cites non self fields &&&&&& \\ \hline
 $MA$ cites non self fields & 0.464 & 0.273 & 0.416 & 0.453 & 0.493 & 0.426 \\ 
 $PHY$ cites non self fields &&&&&& \\ \hline \hline
 $CS$ cites $PHY$ &0.482 & - & 0.187 & 0.092 & 0.30 & 0.371 \\
 $CS$ cites $MA$ &&&&&& \\ \hline
 $PHY$ cites $CS$ &0.038 &    0.155 &    0.090 &    0.002 &    0.009 &    0.032\\
 $PHY$ cites $MA$ &&&&&& \\ \hline
 $MA$ cites $CS$ & 0.023 &    0.073 &    0.041 &    0.008 &    0.010 &    0.014\\ 
 $MA$ cites $PHY$ &&&&&& \\ \hline
\end{tabular}
%\end{adjustbox}
\end{table}
\fi
%\end{adjustbox}
%\end{wraptable}
%\vspace{-0.2cm}
%In order to measure the (dis-)similarity between citing patterns of two fields, we conduct student's $t$-test. Our null hypothesis is that ``citation trends of two different fields are similar''. We analyze citing patterns of fields in three different ways: (a) self-field citing pattern, (b) non self-field citing pattern, and (c) citing pattern toward a particular field. Table~\ref{tab:sig_test} presents $p$-values when two fields are compared against the self-field citations trends. We observe that self-field citation patterns between $CS$, $MA$, and $PHY$ are significantly different. In the majority of cases, we find $p$-values less than 0.05. We also compare non self-field citation patterns of the fields against other fields. Citation patterns of $CS$ and $MA$ and $CS$ and $PHY$ are significantly different. Lastly, we conduct more nuanced experiments by considering pairwise fields. $PHY$ cites 
%$CS$ and $MA$ in a significantly different manner. Similarly, $MA$ cites $PHY$ and $CS$ in a significantly different manner. 

Next, we identify top 12 subfields of each individual field that received the highest number of citations between 1995--2017 (the supplementary material\footnote{\url{https://tinyurl.com/yxrhvufy}} notes the different subfields). The popularity of subfields seems to be inconsistent at different time-periods. It is observed that several subfields have become obsolete over the time with a drastic decrease in their incoming citations. For example, {\em Computation and Language}, a subfield of $CS$, was among top-three most cited subfields during earlier time-periods (\textbf{B1}--\textbf{B3}), but its popularity drastically reduced during the later time-periods (\textbf{B4}--\textbf{B5}). In $CS$, we found no subfield that always exists in the most cited (top three) list. In case of $PHY$, two subfields are always in the most cited list: (i) {\em High Energy Physics - Theory} and (ii) {\em High Energy Physics - Phenomenology}. $MA$ witnesses new subfields such as {\em Group Theory} and {\em Representation Theory}  gaining high popularity whereas old subfields like {\em Logic} and {\em Classical Analysis and ODEs} depleting over the time.
%$CS$'s subfields like \textit{Information Theory}, \textit{Computer Vision and Pattern Recognition}, \textit{Data Structures and Algorithms}, etc. receive majority of citations. Similarly, \textit{Probability}, \textit{Algebraic Geometry},  \textit{Analysis of PDEs}, etc. from $MA$ receive highest number of citations. \textit{Quantum Physics}, \textit{High Energy Physics-Theory}, \textit{General Relativity and Quantum Cosmology}, etc., are most cited in $PHY$. 
\if{0}
\begin{table}[t]
\centering
\caption{Top 12 cited subfields of $CS$, $MA$ and $PHY$.}\label{tab:top_12_subfields}
%\begin{adjustbox}{width=0.8\textwidth}
\tiny
\begin{tabular}{|c|c|c|c|} \hline
\multirow{1}{*}{\bf Ranks} & \textbf{$CS$}&\textbf{$MA$}&\textbf{$PHY$} \\\hline
1&Inf. Theory&Probability&Quant. Phys.\\\hline
2&Comp. Vis. \& Pat. Recog.&Alg Geom. &High Energ. Phys. - Theory\\\hline
3&Data Str. \& Algo.&Analysis of PDEs& Gen. Rel. \& Quant. Cosmo.\\\hline
4& Learning&Combinatorics&High Energ. Phys. - Phen.\\\hline
5&Comp. \& Lang. &Diff. Geom.&Mesoscale \& Nanoscale \\\hline
6&Comp. Complexity&Dyn. Sys.&Stat.  Mech.\\\hline
7&AI&Geom. Topo. &Astrophysics\\\hline
8&Logic in CS&Numb. Theo.&Strongly Corr. Electr.\\\hline
9&CS \& Game Theory &Rep. Theory&Cosmo. \& Nongalactic Astrophys. \\\hline
10& Soc. \& Inf. Net.&Quant. Alg.&Math. Phys \\\hline
11&Discr. Maths&Num. Anal.&Soft Cond. Matt.\\\hline
12&Net. \& Int. Archi. &Grp. Theo. &High Energ. Astrophys. Phen.\\\hline
\end{tabular}
\vspace{-0.5cm}
\end{table}
\fi

\begin{figure*}[t]
\vspace{-0.5cm}
\centering
\setlength\tabcolsep{-1pt}
\begin{tabular}{@{}c@{}c@{}c@{}}
  \includegraphics[width=.33\hsize]{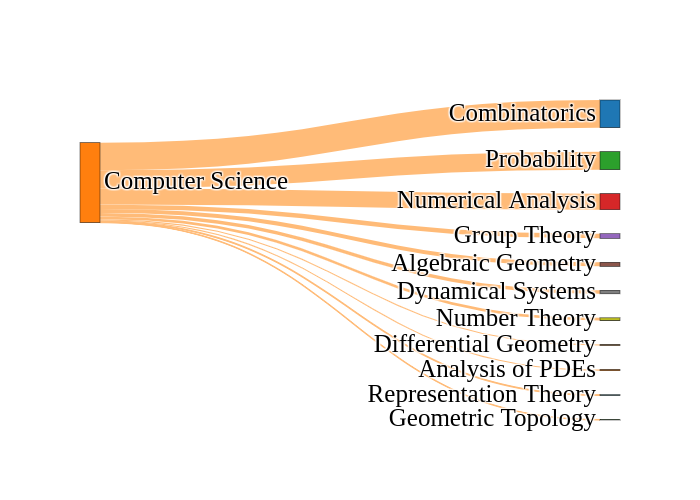} &
  \includegraphics[width=.33\hsize]{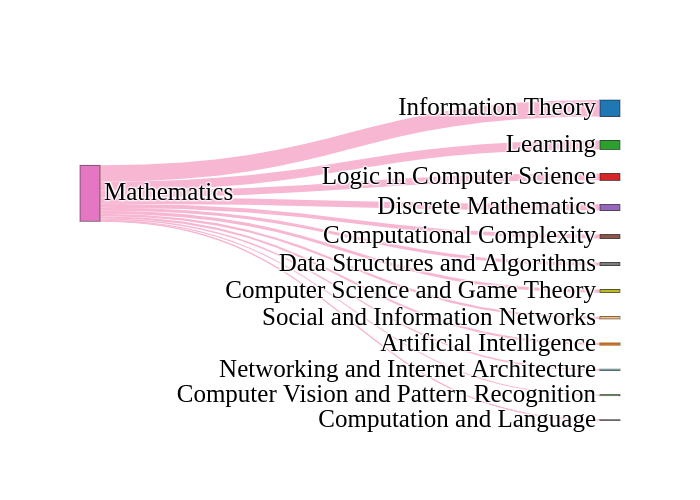}& 
  \includegraphics[width=.33\hsize]{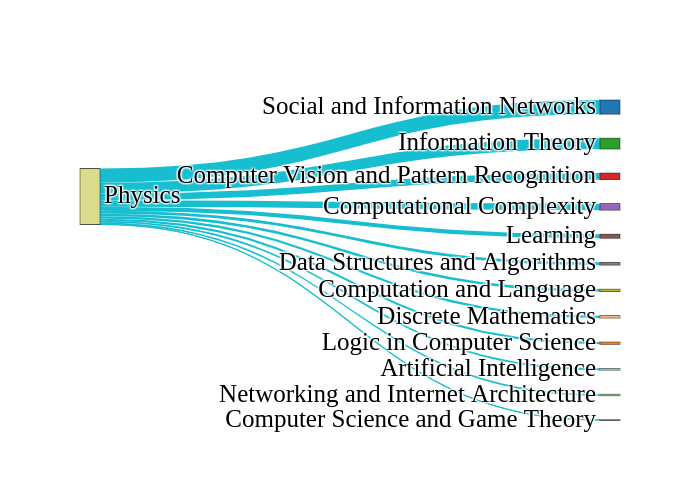}\\
  (a) $CS$ citing $MA$ subfields  & (b) $MA$ citing $CS$ subfields  & (c) $PHY$ citing $CS$ subfields\\
  \includegraphics[width=.33\hsize]{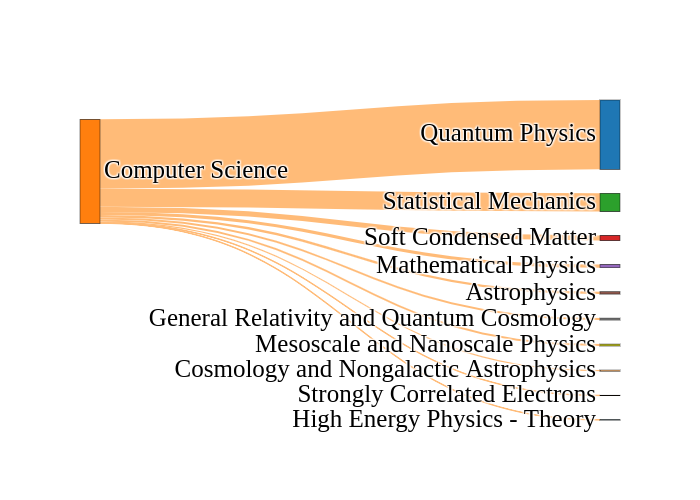} &
  \includegraphics[width=.33\hsize]{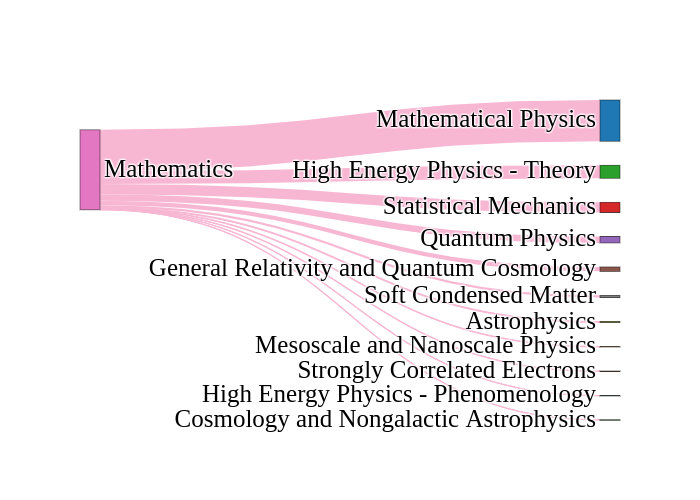}& 
  \includegraphics[width=.33\hsize]{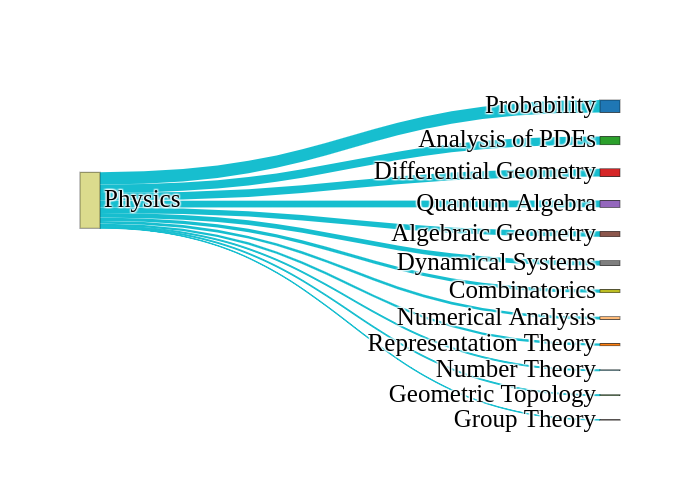}\\
  (d) $CS$ citing $PHY$ subfields & (e) $MA$ citing $PHY$ subfields  & (f) $PHY$ citing $MA$ subfields\\
\end{tabular}
\caption{Citation outflow from fields to other field's subfields over the year range 1995--2017.} 
\label{fig:combined_PHY_MATH}

\end{figure*}

\noindent\textbf{Citation flow analysis}: Next, we perform an analysis of citation flow from fields to subfields. We, again, leverage Sankey diagrams for graphic illustration of the flows. We conduct empirical analysis for the entire time-period along with different temporal buckets. Figure~\ref{fig:combined_PHY_MATH} shows citation flow from each field to other field's subfields over the entire time period. $CS$ mostly cites subfields such as {\em Combinatorics}, {\em Probability}, and {\em Numerical Analysis} from $MA$ and {\em Quantum Physics} and {\em Statistical Mechanics} from $PHY$. Citation inflow from $CS$ to {\em Quantum Physics} is significantly larger than to any other subfield of $PHY$. Similarly, $MA$ mostly cites $CS$ subfields such as {\em Information theory} and  {\em Learning} and $PHY$ subfields such as {\em Mathematical Physics} and {\em High Energy Physics - Theory}. In particular, $MA$ cites {\em Mathematical Physics} in a significantly high proportion ($\sim$51.48\%). $PHY$ mainly cites $CS$ subfields like {\em Social and Information Networks}, {\em Information Theory}, {\em Computer Vision \& Pattern Recognition}, {\em Computational Complexity} and {\em Learning} and $MA$ subfields such as {\em Probability} and {\em Analysis of PDE}. The subfield {\em Information Theory} in $CS$ remains popular for both $PHY$ and $MA$. Similarly, $MA$'s subfield {\em Probability} remains popular for both $CS$ and $PHY$. The most interesting outcome here is the growing interest of $PHY$ and $MA$ in the $CS$ subfield \textit{Learning}. This possibly indicates that in recent times mathematicians and physicists have started taking interest in formulating the mathematical foundations of various machine learning techniques (e.g., theoretical foundations of the deep learning machinery)\footnote{Please see supplementary material: \url{https://tinyurl.com/yxrhvufy}, for more results.}.
\if{0}
%\vspace{-0.9cm}
\begin{figure*}[t]
\centering
\setlength\tabcolsep{-1pt}
\begin{tabular}{@{}c@{}c@{}c@{}}
  \includegraphics[width=.33\hsize]{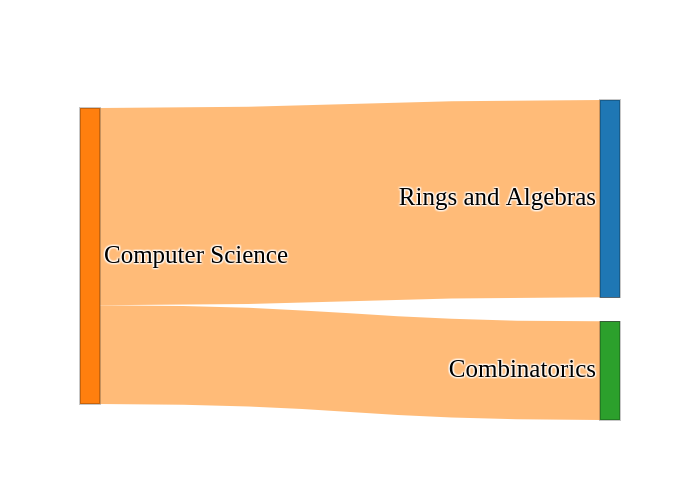} &
  \includegraphics[width=.33\hsize]{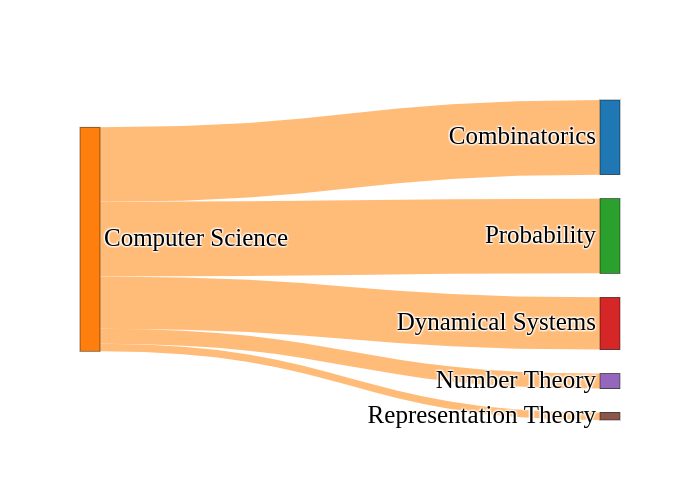}& 
  \includegraphics[width=.33\hsize]{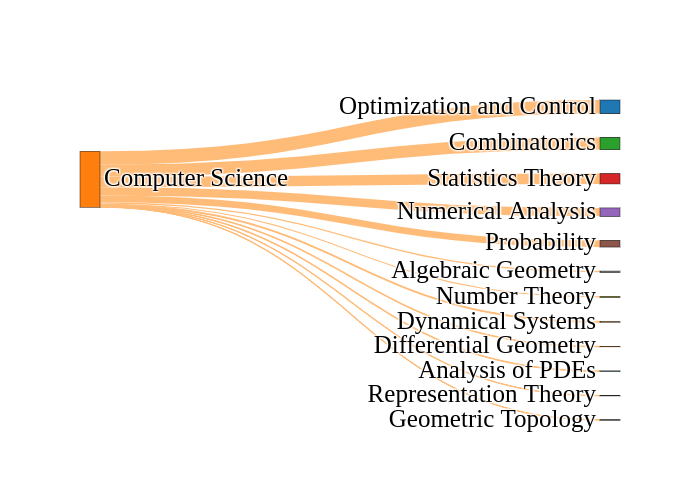}\\
   B3: 2005-2009& B4: 2010-2014 & B5: 2015-2017\\
  & (a) $CS$ to $MA$ &\\
  \includegraphics[width=.33\hsize]{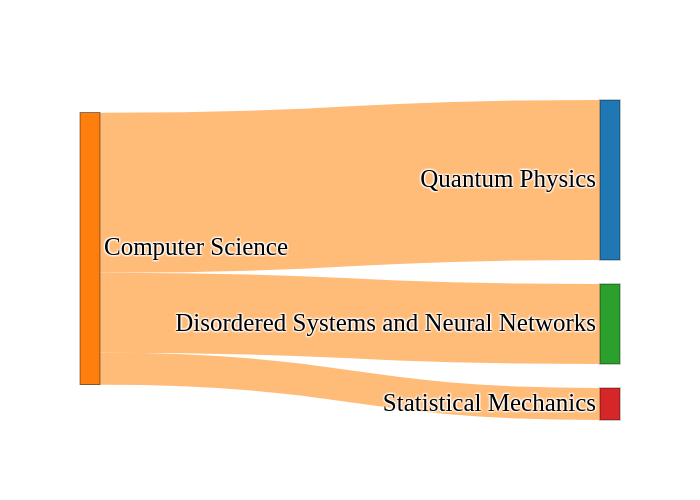} &
  \includegraphics[width=.33\hsize]{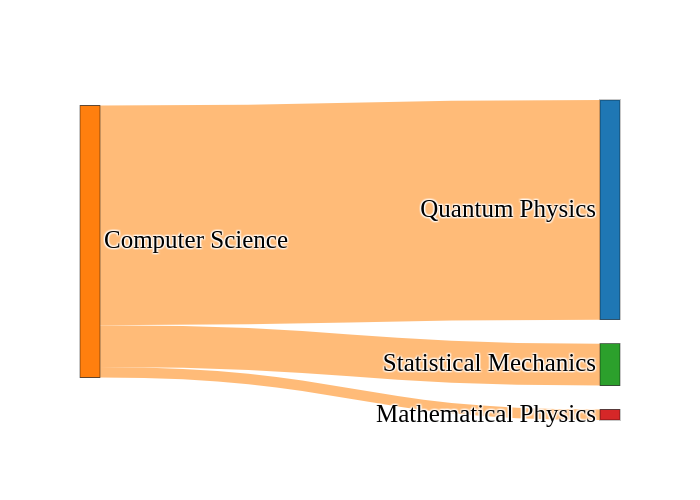}& 
  \includegraphics[width=.33\hsize]{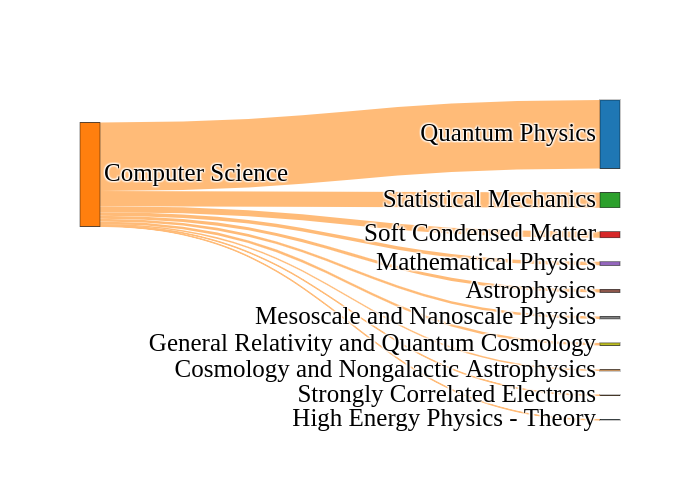}\\
   B3: 2005-2009& B4: 2010-2014 & B5: 2015-2017\\
  & (b) $CS$ to $PHY$ &\\
 
\end{tabular}
\caption{Citation inflow from $CS$ to $MA$ and $PHY$ over three time periods.}% (a) Starting from \textbf{B3}, $CS$ continuously cites large number of articles from {\em Combinatorics}. {\em Probability} entered and gained second position in \textbf{B4} but looses to other sub fields in \textbf{B5}. In the most recent bucket \textbf{B5}, {\em Optimization and Control} tops the chart. (b) {\em Quantum Physics} remains at top for all of the three buckets. {\em Statistical Mechanics} also remains in top three for all the three buckets.}  
\label{fig:combined_CS_MATH_PHY}
\vspace{-0.5cm}
\end{figure*}

Figure~\ref{fig:combined_CS_MATH_PHY}a shows temporal citations inflow from  $CS$ to subfields of $MA$ and $PHY$. Starting from \textbf{B3}, $CS$ continuously cites large number of articles from {\em Combinatorics}. The subfield {\em Probability} entered and gained second position in \textbf{B4} but looses to other subfields in \textbf{B5}. In the most recent bucket \textbf{B5}, {\em Optimization and Control} tops the chart. In case of $CS$ citing subfields of $PHY$ (see figure ~\ref{fig:combined_CS_MATH_PHY}b), {\em Quantum Physics} remains at the top for all of the three buckets. {\em Statistical Mechanics} also remains within the top three for all the three buckets. 

\begin{figure*}[t]
\centering
\setlength\tabcolsep{-1pt}
\begin{tabular}{@{}c@{}c@{}c@{}}
  \includegraphics[width=.33\hsize]{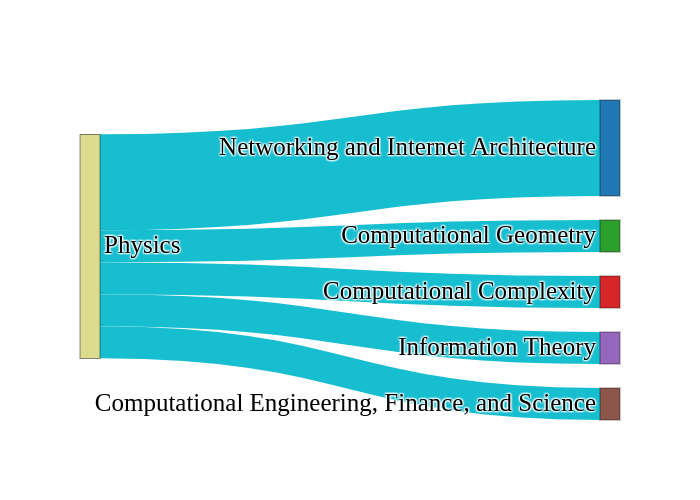}&
  \includegraphics[width=.33\hsize]{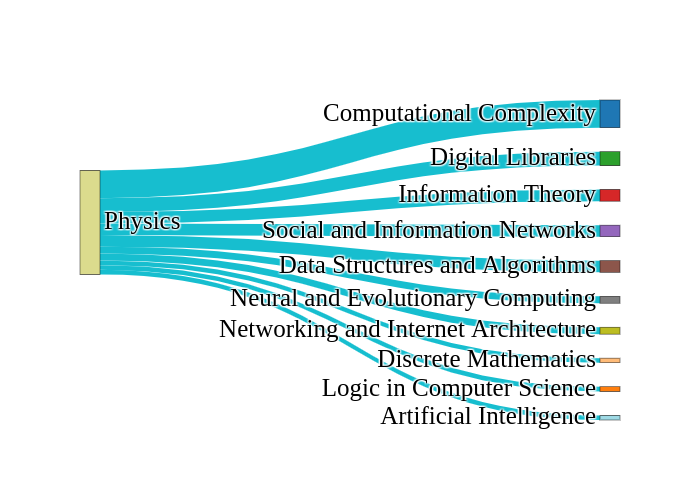}&
  \includegraphics[width=.33\hsize]{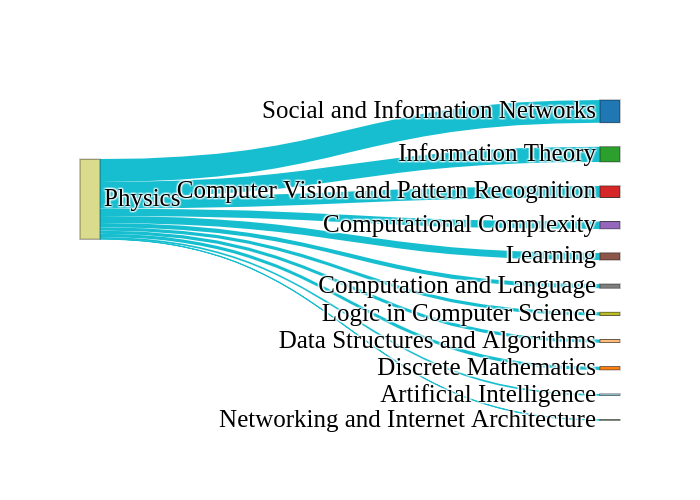}\\
 (a) B3: 2005-2009 & (c) B4: 2010-2014& (d) B5: 2015-2017\\
 & (a) $PHY$ to $CS$ &\\
 \includegraphics[width=.33\hsize]{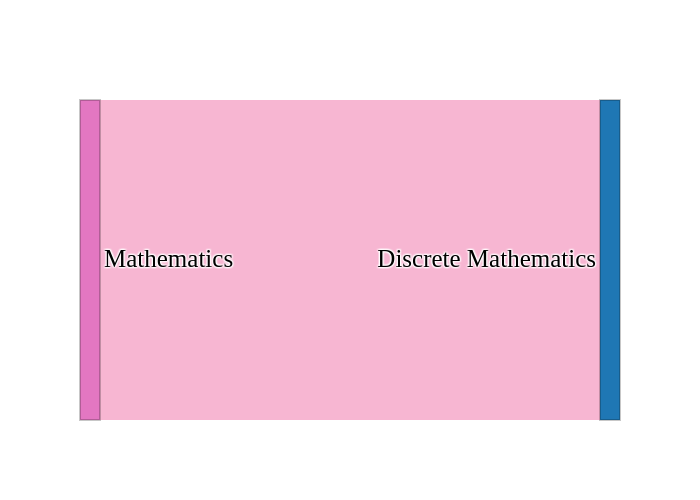} &
  \includegraphics[width=.33\hsize]{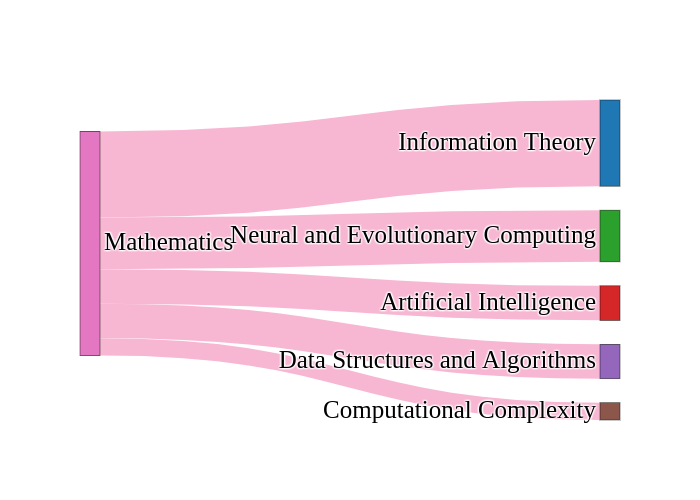}& 
  \includegraphics[width=.33\hsize]{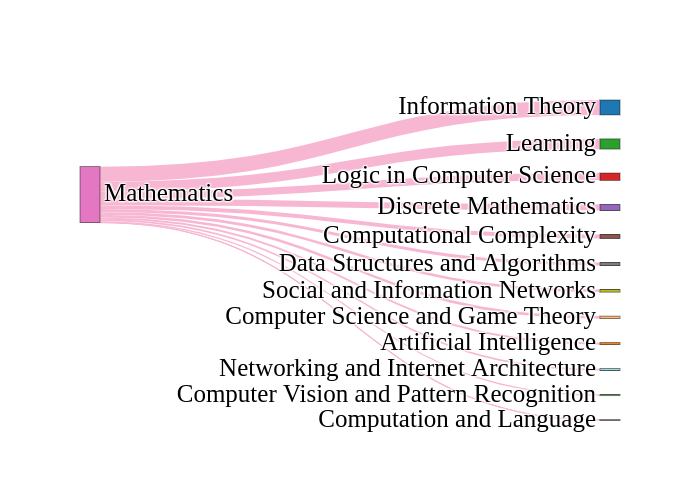}\\
  (a) B3: 2005-2009& (b) B4: 2010-2014 & (c) B5: 2015-2017\\
  & (b) $MA$ to $CS$ &\\
\end{tabular}
\caption{Citation inflow from $PHY$ and $MA$ to $CS$ over three time periods.}
\vspace{-0.5cm}
\label{fig:combined_PHY_MA_CS}

\end{figure*}

\begin{figure*}[t]
\centering
\setlength\tabcolsep{-1pt}
\begin{tabular}{@{}c@{}c@{}c@{}}
  \includegraphics[width=.33\hsize]{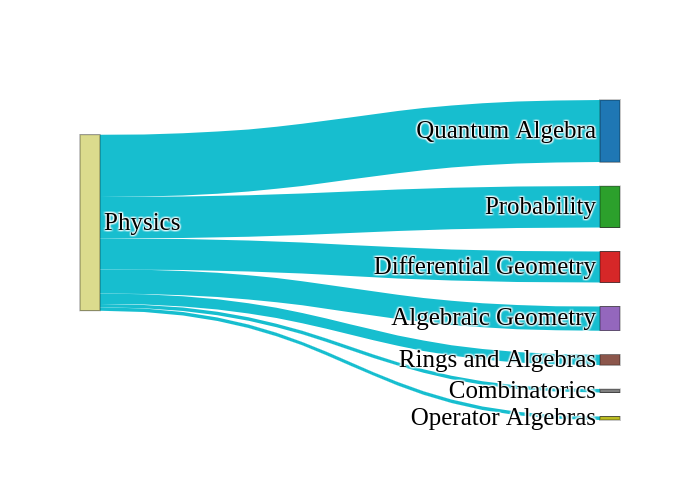}&
  \includegraphics[width=.33\hsize]{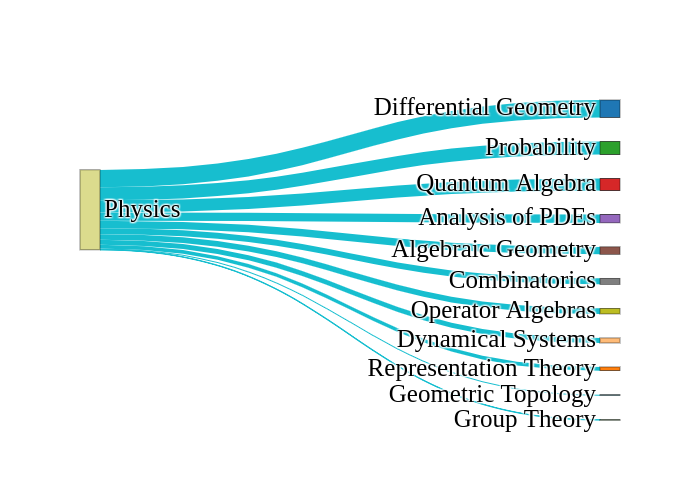}&
  \includegraphics[width=.33\hsize]{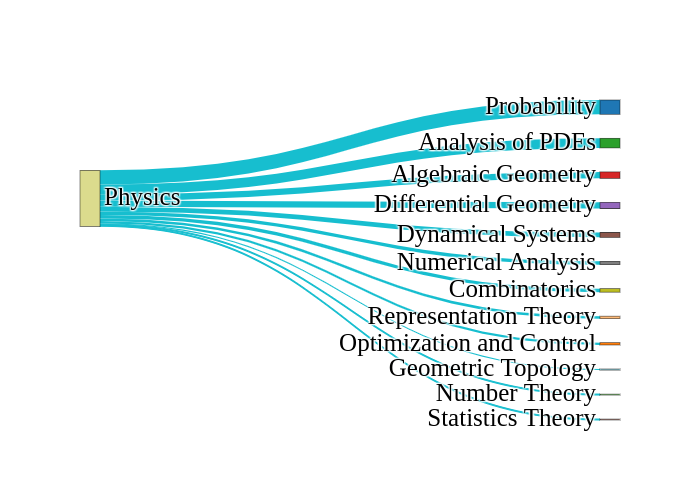}\\
 (a) B3: 2005-2009 & (c) B4: 2010-2014& (d) B5: 2015-2017\\
 & (a) $PHY$ to $MA$ &\\
 \includegraphics[width=.33\hsize]{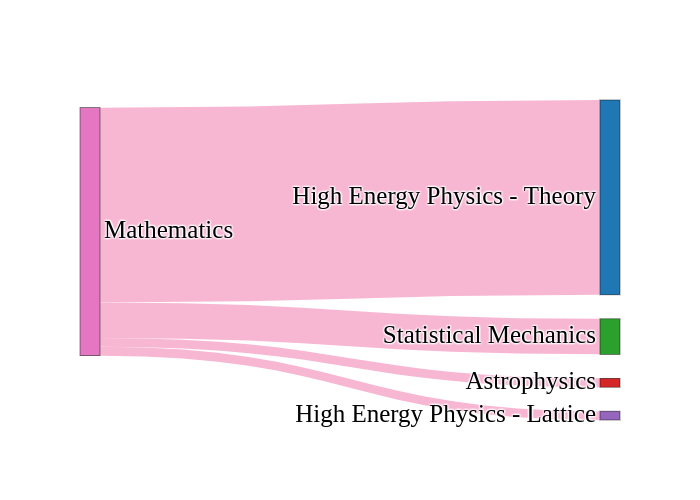} &
  \includegraphics[width=.33\hsize]{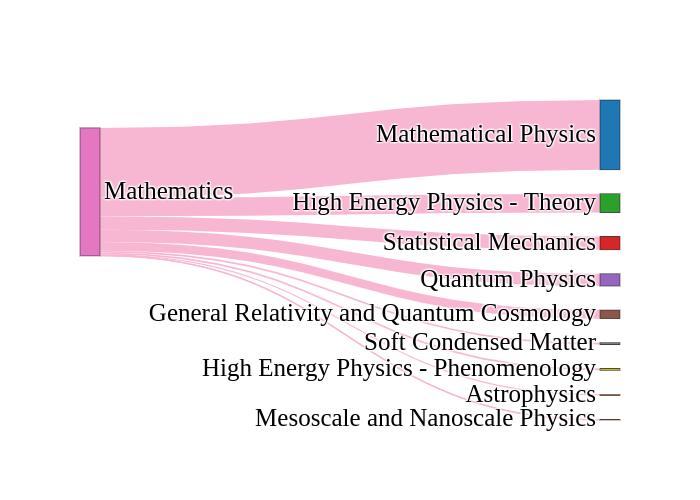}& 
  \includegraphics[width=.33\hsize]{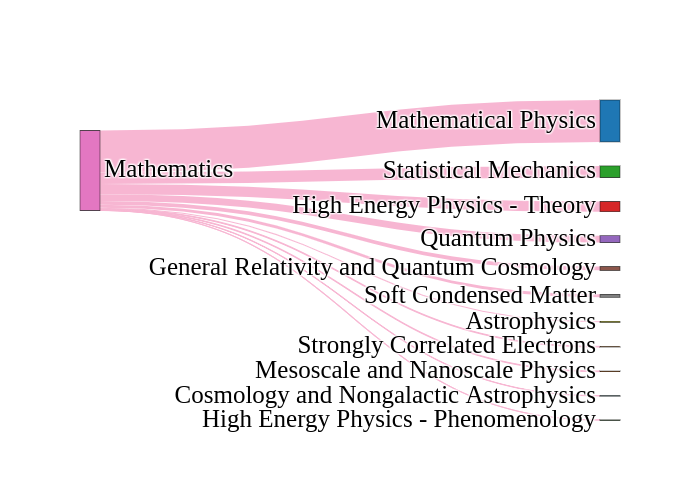}\\
  (a) B3: 2005-2009& (b) B4: 2010-2014 & (c) B5: 2015-2017\\
  & (b) $MA$ to $PHY$ &\\
\end{tabular}
\caption{Citation inflow from $PHY$ to $MA$ and vice-versa over three time periods.}% (a) During initial time period \textbf{B3}, $PHY$ cited large number of papers from {\em Quantum Algebra} papers of $MA$. Later, {\em Differential Geometry and Probability} overcame the popularity of {\em Quantum Algebra} and topped the chart in \textbf{B4} and \textbf{B5} respectively. (b) During \textbf{B3}, {\em High Energy Physics} of $PHY$ tops the chart. {\em Mathematical Physics} started gaining citations in later buckets and tops the chart in \textbf{B4} and \textbf{B5}. {\em Statistical Physics} also received a high volume of citations from $MA$ in later time periods.} 
\label{fig:combined_PHY_MA_PHY}
\vspace{-0.5cm}
\end{figure*}

Figure~\ref{fig:combined_PHY_MA_CS}a
shows citation inflow from $PHY$ to $CS$ over three temporal buckets. During \textbf{B3}, it only cites {\em Discrete Mathematics}. However, in the next bucket \textbf{B4}, {\em Information Theory} tops the chart. {\em Neural \& Evolutionary Computing Mathematics} and  {\em Artificial Intelligence} came at second and third positions, respectively. {\em Information Theory} continues to top the chart in \textbf{B5} too.  Figure ~\ref{fig:combined_PHY_MA_CS}b shows that during initial time-period, $PHY$ highly cites {\em Information Theory} papers of $CS$. 

Figure~\ref{fig:combined_PHY_MA_PHY}a shows citation inflow from $PHY$ to $MA$ over three temporal buckets. During initial times, $PHY$ cited large number of papers from the {\em Quantum Algebra} sub field of $MA$. Later, {\em Differential Geometry and Probability} overcame the popularity of {\em Quantum Algebra} and topped the chart in \textbf{B4} and \textbf{B5} respectively. Figure~\ref{fig:combined_PHY_MA_PHY}b shows citation inflow from $MA$ to $PHY$ over three temporal buckets. During \textbf{B3}, {\em High Energy Physics} of $PHY$ is mostly cited. {\em Mathematical Physics} started gaining citations in later buckets and tops the chart in \textbf{B4} and \textbf{B5}. \fi

%\vspace{-0.2cm}

\section{Conclusion and Future Work}
\label{sec:end}
We study a large collection of research articles from three different scientific disciplines -- \textit{PH}, \textit{MA}, and \textit{CS} to understand how citation patterns from the core science fields to applied field have drastically changed. Besides, our work raises some fundamental questions such as which factors of a subfield are responsible for gaining citations from other disciplines? Is it related to the development of that subfield by borrowing ideas from other disciplines or due to the appearance of a new idea in that subfield that attracts attention from other disciplines? This can be studied by identifying seminal contributions in that subfield and its relation to the significant ideas of other subfields which are cited by the original subfield. 

\bibliographystyle{splncs04}
\bibliography{main}

\end{document}